\documentclass[]{JHEP3}
\title{The Fuzzy Sphere:From The Uncertainty Relation To The
Stereographic Projection.}
\author{Musongela Lubo\\The Abdus Salam International Centre for Theoretical Physics 
P.O.Box 586\\ 34100 Trieste, Italy.\\ E-mail:muso@ictp.trieste.it}
\abstract{On the fuzzy sphere, no state saturates simultaneously all the
Heisenberg uncertainties. We propose a weaker uncertainty for which this holds.
The family of states so obtained is physically 
motivated because it encodes information about positions in this fuzzy
context. In particular, these states realize in a natural way a deformation
of the stereographic
projection. Surprisingly, in the large $j$ limit, they reproduce some properties of 
the ordinary coherent states on the non commutative plane.}
\keywords{fuzzy sphere, coherent states, star product}
\preprint{}
\dedicated{}
\begin{document}

\section{Introduction.}

The idea that at very small scale the physical structure of space time may display
non trivial features has received wide interest. In particular non commutative
theories have been under scrutiny in the last years. One of the key tools
in this framework is the star product, which allows the treatment of these
theories using commuting variables.
The first star product concerned  phase space \cite{moyal1,moyal2,voros}; it is 
readily transfered to the non commutative plane composed of two position
operators whose commutator is a constant.   

Among the methods used to build  star products, the one based on coherent states
is especially interesting since associativity is then automatically
assured. This is  a consequence of the fact that  coherent states
form an over complete basis on the Hilbert space on which the algebra of 
quantum operators act.

The terms "coherent states" and "generalized coherent states" are used in the literature
following the works of Perelomov and Berezin for example. These notions are defined in
connection with irreducible representations of Lie groups and quotient spaces by these 
groups. The states appearing here arise differently: we generalize a different property
of the system of coherent states on the non commutative plane. 

On the non commutative plane, coherent states are obtained in a
straightforward way. They saturate the bound of the Heisenberg uncertainty
and this is equivalent to the fact that they are eigenstates of the
destruction operator \cite{pere,klau}. On the fuzzy sphere, the situation is quite 
different. Firstly, there is no state saturating all the uncertainties. Secondly,
the identification of a destruction operator is not straightforward;
one possibility is the use of the stereographic projection to obtain a deformed
creation-destruction algebra.
The problem then
is the difference between the dimensions of the Fock spaces;
one usually wishes to use finite dimensional representations of $SU(2)$ for the fuzzy
sphere but
the Fock space on which the deformed creation-destruction algebra acts is
infinite dimensional. One of the proposals to tackle this problem was trough a
truncature procedure \cite{pinzul1}. 

We propose here another generalization which starting point is not a deformed
destruction operator but an enlarged Heisenberg uncertainty. The states 
saturating the associated bounds are called "generalized" squeezed states. 
Another approach, relying on a functional \cite{lubo1} and exploiting an idea
introduced in another context \cite{spindel}, led to  technical problems.
The procedure followed here is not as general and ambitious as  the ones developed in 
\cite{kon,za,pre1,pre2}. Like in \cite{pinzul1} and contrary to \cite{pre1,pre2},
we expect the induced star product to be expressed in terms of two independent rather than three
dependent coordinates. The derivation and  the use of the associated star 
product to study quantum field theory like in \cite{wata1,wata2,wata3,
bala1,bala2,bala3,ho,star1,star2,star3,star4,star5,star6} and especially the Seiberg-Witten map \cite{seiberg} are
our final goal \cite{lubo2}.   

This paper is organized as follows. In the second section we summarize the
construction of coherent states on the fuzzy sphere developed in \cite{pinzul1} 
and discuss its differences with the procedure used for  the non commutative plane. The third
section is devoted to the derivation of what we call generalized squeezed 
states  on a two dimensional space. In section $4$ we show that while the
Heisenberg uncertainties on the non commutative sphere
can not be saturated simultaneously, the ones we 
derived in the previous section just do that. We obtain that the scalar product
of two  generalized squeezed states reproduces, in the $j\rightarrow \infty$ limit,
the corresponding
expression on a non commutative plane. In section $5$, we show that the mean 
values of
the position operators on these states lead to the stereographic projection.

We emphasize that the appearance of the stereographic projection obtained
here is of
a different nature compared to the ones of \cite{pinzul1,arik}. In these two
works, the stereographic projection is introduced from the start to define
a pair of  creation-destruction like operators from the three quantum
ones defining the fuzzy sphere. On the contrary, we begin with an
uncertainty relation which is weaker than the Heisenberg's and find that
it leads to a family of states parameterized by a complex number in a way which
 brings in the stereographic map.

\section{Coherent States: \newline The Non Commutative Plane  Versus  The Fuzzy Sphere.}
   
The non commutative plane is defined by the  relations
\cite{nekra}
\begin{equation}
\label{eqb1}
[ \hat{x}_j , \hat{x}_k ] = i \theta \, \epsilon_{j,k}  \quad , \quad \theta > 0 
\quad , \quad k,j=1,2 \quad .
\end{equation}
The coherent states are defined as eigenvalues of a destruction operator
and saturate the Heisenberg uncertainty $\Delta x_1 \Delta x_2 = \theta/2 . $

The situation of the fuzzy sphere is far from being so simple. 
It is  a matrix model defined by  
the following relations \cite{madore1,madore2}:
\begin{eqnarray}
\label{eqb2}
[ \hat x_k , \hat x_l ] &=& \frac{i R}{\sqrt{j(j+1)}} \epsilon_{klm} \hat x_m \quad , \quad
\delta^{l k} \hat{x}_l \hat{x}_k = R^2  \quad , \nonumber\\
& & {\rm with}  
\quad j \quad  {\rm integer} \quad {\rm or } \quad {\rm half-integer} \quad
{\rm and} \quad k,l,m =1,2,3 \quad .
\end{eqnarray}
We will use rescaled variables such that the radius $R$ can be given the value
one. There is no state saturating simultaneously all the Heisenberg uncertainties
\cite{lubo1}
\begin{eqnarray}
\label{eqb3}
\Delta x_1 \Delta x_2 &=& \frac{1}{2} \frac{1}{\sqrt{j(j+1)}} \vert \langle \hat{x}_3 \rangle \vert \quad , \quad 
\Delta x_2 \Delta x_3 = \frac{1}{2} \frac{1}{\sqrt{j(j+1)}} 
\vert \langle \hat{x}_1 \rangle \vert \quad , \nonumber\\
\Delta x_3 \Delta x_1 &=& \frac{1}{2} \frac{1}{\sqrt{j(j+1)}} \vert \langle \hat{x}_2 \rangle \vert \quad ;
\end{eqnarray}
we will come back to this point in the fourth section. As a consequence, one has to 
resort to other criteria to define  coherent states
in this context.  
One way to tackle this problem is based on the construction of a deformation of the 
creation-destruction operators \cite{pinzul1}. This is summarized below. 

One uses the
stereographic projection to define the operators
\begin{equation}
\label{eqb4}
z = ( \hat{x}_1 - i \hat{x}_2) (1 - \hat{x}_3 )^{-1} \quad , \quad
z^+ = (1 - \hat{x}_3 )^{-1} ( \hat{x}_1 + i \hat{x}_2) 
\end{equation}
which obey the commutation relation
\begin{equation}
\label{eqb5}
[ z , z^+ ] = F(z z^+) \quad ,
\end{equation}
where
\begin{eqnarray}
\label{eqb6}
F(z z^+) &=& \alpha \chi  \left[ 1 + \vert z \vert^2 - 
    \frac{1}{2} \chi  \left(1 + \frac{\alpha}{2} \vert z \vert^2 \right) \right]
    \quad , \quad
\chi = \frac{2}{\alpha} \left[ 1 + \frac{\alpha}{2 \xi} -   
\sqrt{\frac{1}{\xi} + \left( \frac{\alpha}{2 \xi} \right)^2} \right] \quad ,\nonumber\\
\xi &=& 1 + \alpha \vert z \vert^2 \quad , \quad {\rm and} \quad \alpha = \frac{1}{\sqrt{j(j+1)}}
\quad .
\end{eqnarray}
Thanks to Eq.(\ref{eqb5}), there exists a function $f$ which relates the 
operators $z,z^+$ to the usual
creation and destruction operators $\hat{a}^+,\hat a$:
\begin{equation}
\label{eqb7}
z = f( \hat{a}^+ \hat{a} + 1) \hat{a} \quad . 
\end{equation}
The generalized coherent states $\vert \zeta \rangle$ are taken to be the eigenstates of the
destruction operator $z$ and read 
\begin{equation}
\label{eqb8}
\vert \zeta \rangle= N(\vert \zeta \vert^2)^{-1/2} \quad 
\exp{ \left[ \zeta f^{-1}(\hat{a}^+ \hat{a} ) \hat{a}^+ \right] } f^{-1}( 
\hat{a}^+ \hat{a}  )
\vert 0 \rangle \quad ,
\end{equation}
the function $ N(\vert \zeta \vert^2)^{-1/2} $ enforcing the normalization of the wave 
function. The vacuum $\vert 0 \rangle$ is annihilated by $\hat a$. 

It should be stressed that the map between the couples of operators $z,z^+$ and
 $\hat{a},\hat{a}^+$ is singular since the second pair act on an infinite Fock space
while  the first  acts on a finite 
dimensional vector space. This was taken into account by restricting, for a fixed $j$, the 
expansion of Eq.(\ref{eqb8}) to the component $\vert 2 j \rangle$ :
\begin{eqnarray}
\label{eqb9}
\vert \zeta , j \rangle &=& N_j(\vert \zeta \vert^2)^{-1/2} 
\sum_{n=0}^{2 j} \frac{\zeta^n}{ \sqrt{n!} [ f_j(n) ]!} \vert n \rangle \quad , \quad
 N_j(x) =  
\sum_{n=0}^{2 j} \frac{x^n}{ n! ([ f_j(n) ]!)^2 } \quad , \nonumber\\
f_j(n) &=&  \frac{\sqrt{2 j - n +1}}{  \sqrt{j(j+1)} + j - n} \quad , \quad
 {\rm where}  \quad
[f(n)]!=f(n) f(n-1) ... f(0) \quad .
\end{eqnarray}
The states obtained in this way are not eigenstates of the operator $z$ anymore.

The complications about coherent states on the fuzzy sphere come from the fact 
that, contrary to the 
non commutative plane, the Heisenberg uncertainties given in Eq.(\ref{eqb3}) 
can not be saturated 
simultaneously. If one relies on a mapping from $z, z^+$ to the ordinary 
creation-destruction 
operator, one has to face  the discrepancy between the dimensions of the 
spaces involved. If one
cuts the sum at a given index, the states obtained are 
not exact eigenstates
of the operator $z$.  

The main point of this work is that there is a natural extension of the Heisenberg uncertainty
which allows one to define  "generalized squeezed states" for the fuzzy sphere in a way
which is more or less similar to what one does for the quantum plane. The
dimensionality problem pointed above is absent in this approach. 

In brief, we shall show
that having two operators $\hat x, \hat p$, the most general uncertainty relation implying the mean
values of expressions at most quadratic in the operators is
\begin{equation}
\label{eqb19}
  (\Delta x)^2  (\Delta p)^2  \geq \frac{1}{4}  
  \left[  (  \langle  \{ \hat x , \hat p  \} \rangle
    - 2
\langle \hat p\rangle  \langle \hat x\rangle )^2 + \vert \langle [ \hat x , \hat p] \rangle  \vert^2 
\right] 
\quad. \end{equation}
This relation involves not only the commutator but also the anti commutator; the
Heisenberg uncertainty can be deduced from it. 
For ordinary coherent states, the first quantity under square brackets on the right
side vanishes 
 and the Heisenberg bound is attained. We shall show that the equations
\begin{eqnarray}
\label{eqb20}
  (\Delta x_1)^2  (\Delta x_2)^2  &=& \frac{1}{4}  
  \left[  (  \langle  \{ \hat{x}_1 , \hat{x}_2  \} \rangle
    - 2
\langle \hat{x}_1 \rangle  \langle \hat{x}_2 \rangle )^2 + 
\vert \langle [ \hat{x}_1 , \hat{x}_2] \rangle  \vert^2 
\right]  \quad ,  \nonumber\\
(\Delta x_2)^2  (\Delta x_3)^2  &=& \frac{1}{4}  
  \left[  (  \langle  \{ \hat{x}_2 , \hat{x}_3  \} \rangle
    - 2
\langle \hat{x}_2 \rangle  \langle \hat{x}_3 \rangle )^2 + 
\vert \langle [ \hat{x}_2 , \hat{x}_3] \rangle  \vert^2 
\right]   \quad , \nonumber\\
(\Delta x_3)^2  (\Delta x_1)^2  &=& \frac{1}{4}  
  \left[  (  \langle  \{ \hat{x}_3 , \hat{x}_1  \} \rangle
    - 2
\langle \hat{x}_3 \rangle  \langle \hat{x}_1 \rangle )^2 + 
\vert \langle [ \hat{x}_3 , \hat{x}_1] \rangle  \vert^2 
\right]  \quad . 
\end{eqnarray}
are compatible on the fuzzy sphere. The states obeying them simultaneously are what
we shall call "generalized squeezed states". We will show that they are
parameterized by a point on the complex plane and possess other interesting
properties.

\section{Squeezed States: The $2D$ Phase Space Revisited.}

To begin, let us consider a two dimensional phase space whose quantum
operators are denoted $\hat x , \hat p$. We will work on  a linear combination 
of the two operators supplemented by a term proportional to the identity: 
\begin{equation}
\label{eqc1}
\hat{\Theta}_\lambda = \hat x + ( \lambda_1 + i \lambda_2) \hat p + ( \lambda_3 + i
\lambda_4) I \quad ,
\end{equation}
the $\lambda_i$ being real constants. 
In the derivation of the Heisenberg
inequality, one assumes for example $ \lambda_1 =  0 $. We will recover the known result as a particular
case. 

For any state $\vert \psi \rangle $ and any  real 
$\lambda_i$, the following quantity is positive or null: 
\begin{equation} 
\label{eqc2} 
\vert \vert  \hat{\Theta}_\lambda  \vert \psi \rangle \vert \vert^2 \geq 0   \quad
.\end{equation}  
If one fixes the state $\vert \psi \rangle $, the preceding function is 
a second degree polynomial in the $\lambda_i$:
\begin{eqnarray}
\label{eqc3}
\varphi(\lambda_1,\lambda_2, \lambda_3, \lambda_4) &=& 
\langle \psi \vert  \hat{\Theta}_\lambda^+ \hat{\Theta}_\lambda  \vert \psi \rangle 
\nonumber\\
&=&
 ( \lambda_1^2 +
\lambda_2^2) \langle {\hat p}^2 \rangle + ( \lambda_3^2 + \lambda_4^2) +
2 ( \lambda_1 \lambda_3 + \lambda_2 \lambda_4) \langle \hat p \rangle 
+ \lambda_1  A_{xp}  \nonumber\\ 
&+&\lambda_2  C_{xp} +  2 \lambda_3  \langle \hat x \rangle + \langle {\hat
x}^2 \rangle \quad .
\end{eqnarray}
We have used the notations 
\begin{equation}
\label{eqc4}
A_{xp} = \langle  \hat x \hat p + \hat p \hat x \rangle \quad ; \quad 
C_{xp} = i ( \langle  \hat x \hat p - \hat p \hat x \rangle ) 
\end{equation} 
to simplify future formulas and assumed a normalization to unity: $\langle \psi \vert \psi\rangle = 1.$
Due to hermiticity conditions, the mean values $A_{xp}$ and $C_{xp}$ are real.
Our polynomial can be diagonalized on its real variables:
\begin{eqnarray}
\label{eqc5}
\varphi(\lambda_1,\lambda_2, \lambda_3, \lambda_4) &=&  \langle {\hat
p}^2   \rangle \left[ {\bf \lambda_1} +  \left(  \frac{1}{2} A_{xp} + 
\langle  \hat p \rangle  {\bf \lambda_3} 
\right) \frac{1}{\langle {\hat p}^2 \rangle}   \right]^2  +
\langle {\hat
p}^2   \rangle \left[ {\bf \lambda_2} +  \left(  \frac{1}{2} C_{xp} + 
\langle  \hat p \rangle  {\bf \lambda_4} 
\right) \frac{1}{\langle {\hat p}^2 \rangle}   \right]^2  \nonumber\\
&+& \frac{(\Delta p)^2}{\langle {\hat
p}^2   \rangle } \left[ {\bf \lambda_3} + \frac{2 \langle \hat x \rangle 
\langle {\hat
p}^2 \rangle - A_{xp} \langle \hat p \rangle}{2 (\Delta p)^2} \right]^2  + \frac{(\Delta
p)^2}{\langle {\hat p}^2   \rangle } \left[ {\bf \lambda_4} - \frac{ C_{xp} \langle \hat p \rangle}{2
(\Delta p)^2} \right]^2  \nonumber\\
&+& \left[ \langle {\hat x}^2   \rangle  - \frac{1}{4} \frac{A^2_{xp}}{\langle
{\hat p}^2   \rangle  } -   \frac{1}{4} \frac{C^2_{xp}}{\langle {\hat p}^2  
\rangle  } - \frac{1}{4}  \frac{(2 \langle \hat x \rangle \langle {\hat p}^2
\rangle - A_{xp} \langle \hat p \rangle)^2}{ \langle {\hat p}^2  
\rangle(\Delta p)^2} - \frac{1}{4} \frac{C^2_{xp} \langle \hat p
\rangle^2}{\langle {\hat p}^2   \rangle(\Delta p)^2} \right]   \quad .
\end{eqnarray}

The content of the four first brackets in this formula form a complete set of 
independent variables built from the $\lambda_i$ so that the positivity of our 
polynomial  implies  $(\Delta p)^2 , \langle
{\hat p}^2   \rangle > 0$ which are trivially verified. The only
valuable inequality we infer is that
\begin{equation}
\label{eqc6}
\min \varphi =   \langle {\hat x}^2   \rangle  - \frac{1}{4}
\frac{A^2_{xp}}{\langle {\hat p}^2   \rangle  } -   \frac{1}{4}
\frac{C^2_{xp}}{\langle {\hat p}^2   \rangle  } - \frac{1}{4}  \frac{(2
\langle \hat x \rangle \langle {\hat p}^2 \rangle - A_{xp} \langle \hat p
\rangle)^2}{ \langle {\hat p}^2   \rangle(\Delta p)^2} - \frac{1}{4}
\frac{C^2_{xp} \langle \hat p \rangle^2}{\langle {\hat p}^2   \rangle(\Delta
p)^2}  \geq 0 \quad ,
 \end{equation} 
since the polynomial is  positive or null for any choice of
the variables $\lambda_i$. This minimum is attained for the following values of
these variables:
\begin{eqnarray}
\label{eqc7}
\bar\lambda_1 &=& - \frac{1}{2 \langle {\hat p}^2 \rangle}
 ( A_{xp} - 2 \langle \hat p \rangle \langle \hat x \rangle ) 
 \left( 1 + \frac{\langle \hat p \rangle^2}{2 (\Delta p)^2} \right) \quad ,
\quad \bar\lambda_2 = -\frac{1}{2} \frac{ C_{xp}}{(\Delta p)^2}  \nonumber\\
\bar\lambda_3 &=& - \langle \hat x \rangle - \frac{\langle \hat{p} \rangle }{2
(\Delta p)^2}  \left( - A_{xp} + 2 \langle \hat{p} \rangle \langle \hat{x} \rangle 
\right)  \quad , \quad  \bar\lambda_4 = \frac{1}{2} \frac{\langle \hat p
\rangle }{(\Delta p)^2}  C_{xp} \quad .
 \end{eqnarray} 

The  inequality displayed in Eq.(\ref{eqc6}) involves
all momenta of first and second degree in the phase space variables. It
contains  the anticommutator and the commutator mean values $A_{xp}$ and
$C_{xp}$. The apparent lack of  symmetry between the position and the momentum
mean values $ \langle \hat x \rangle , \langle \hat p \rangle $ is linked to
the fact that the operator of Eq.(\ref{eqc1}) was not symmetric in the two 
operators; moreover we broke the symmetry between  the $ \lambda_i $  in our
diagonalization. 

Actually, this inequality can be rewritten in a closer and more symmetric form.
First, let us consider its second and fourth terms which are the only ones containing
the anti commutator. Putting them on the same denominator, one obtains
\begin{eqnarray}
\label{ajout1}
&-& \frac{1}{4 \langle \hat{p}^2 \rangle (\Delta p)^2} 
\left[ (  (\Delta p)^2 + \langle \hat p \rangle^2 ) A_{xp}^2 - 2 
\langle \hat x \rangle \langle \hat{p}^2 \rangle  \langle \hat p\rangle A_{xp}
+ 4 \langle \hat x \rangle^2 \langle \hat p^2 \rangle^2 \right] \nonumber\\
&=& \frac{1}{4 (\Delta p)^2} 
\left[( A_{xp} - 2 \langle \hat x \rangle \langle \hat p \rangle )^2 + 
4 \langle \hat x \rangle^2 (\Delta p)^2 \right] \quad .
\end{eqnarray}
In the same way, the third and fifth terms which contain the anticommutator can be
rewritten as
\begin{eqnarray}
\label{ajout2}
\frac{1}{4 (\Delta p)^2} C_{xp}^2 \quad .
\end{eqnarray}
The inequality obtained above can now be rewritten in
the more compact form 
 \begin{equation}
\label{eqc8}
  (\Delta x)^2  (\Delta p)^2  \geq \frac{1}{4}  \left(  ( A_{xp} - 2
\langle \hat p\rangle  \langle \hat x\rangle )^2 + C^2_{xp}  \right) 
\quad. \end{equation} 

The minimum of the function $\varphi$ solely depends on the state $\vert \psi
\rangle$ as can readily be seen in (\ref{eqc6}). We can go one step further and
ask  which states make the minimum of $\varphi$ attain its lowest i.e zero
value. Such a state will obviously satisfy  the equation
\begin{equation}
\label{eqc9}
\hat{\Theta}_{\bar\lambda}  \vert \psi \rangle = 0  
\end{equation} 
and the equality will be attained in Eq.(\ref{eqc8}), the parameters  $\bar\lambda_i$
being related to the mean values of the state by Eqs.(\ref{eqc7}).

From this we can derive the Heisenberg uncertainty
relation 
\begin{equation}
\label{eqc10}
\Delta x \Delta p \geq \frac{1}{2}  \vert C_{xp} \vert  \quad ;
\end{equation} 
when  it is saturated, the following equality holds
\begin{equation}
\label{eqc11}
 \tilde{A}_{xp} = A_{xp} - 2
\langle \hat p\rangle  \langle \hat x\rangle  = \langle \hat x \hat p + \hat
p \hat x \rangle - 2 \langle \hat p\rangle  \langle \hat x\rangle  = 0 \quad .
\end{equation} 
This relation can be used to eliminate $A_{xp}$ in Eq.(\ref{eqc7}); one then obtains the simpler
expressions
 \begin{equation}
\label{eqc12}
\bar{\lambda}_1 = 0 \quad , \quad \bar{\lambda}_3 = - \langle \hat x \rangle \quad ;
\end{equation} 
 Eq.(\ref{eqc9}) then assumes the well known form
\begin{equation}
\label{eqc13}
\left( \hat x - \langle \hat x \rangle + \frac{C_{xp}}{2 (\Delta p)^2}  ( \hat
p - \langle \hat p \rangle )  \right) \vert \psi \rangle = 0  \quad . 
\end{equation}

 The main result here is  given in Eq.(\ref{eqc8}) and the conclusion
that 
\begin{equation}
\label{eqc14}
\Delta x \Delta p = \frac{1}{2}  \vert \langle \hat x \hat p - \hat p \hat x
\rangle \vert  \Longleftrightarrow \langle \hat x \hat p + \hat p \hat x
\rangle - 2 \langle \hat p\rangle  \langle \hat x\rangle  = 0  \quad . \end{equation} 
This statement can be verified in the ordinary theory. The commutation 
relation is then given by $
[ \hat x , \hat p ] = i  $ .
Working in the momentum representation, the solution of Eq.(\ref{eqc13}) has
the well known Gaussian form  \cite{cohen}
\begin{equation}
\label{eqc15}
\psi = \left[ \exp{\left( -\frac{b_1^2}{2 a} \right)} \sqrt{- \frac{\pi}{2 a}} 
\right]^{-1/2} 
e^{(a p^2+(b_1 + i b_2) p)} \quad   \quad .
\end{equation} 
One has
\begin{eqnarray}
 \langle \hat x \rangle = - b_2 \quad , \quad 
 \langle \hat p \rangle = - \frac{b_1}{2 a} \quad , \quad
  \langle \hat p \hat x \rangle = - \frac{i}{2} +  \frac{b_1 b_2}{2 a} \quad , \quad
  \langle \hat x \hat p \rangle =  \frac{i}{2} +  \frac{b_1 b_2}{2 a} \quad , \quad
\end{eqnarray} 
so that Eq.(\ref{eqc14}) is verified.
The situation is the same in the one dimensional K.M.M \cite{kmm1}
model which is defined
by the commutation rules 
$ [ \hat x, \hat p ] = i  ( 1 +  \hat{p}^2) \quad . $
A momentum representation exists in which the action of the operators and the
scalar product defining the Hilbert space are given by    
\begin{equation}
\label{eqc17}
\hat x = i ( 1 + p^2 ) \partial_p  \quad , \quad  \hat p = p \quad , \quad
\langle  \phi \vert \psi \rangle  =  \int  dp \frac{1}{1+p^2}  \phi^*(p)
\psi(p) \quad . \end{equation} 
This theory admits a minimal uncertainty in length: $\Delta x \geq 1$.
Let us verify the left part of Eq.(\ref{eqc14}) for the most important 
solutions  of
Eq.(\ref{eqc13}) in this theory i.e those which exhibit  the minimal uncertainty
in length:  
\begin{equation} 
\label{eqc18}
\psi(p) = \sqrt{\frac{2}{\pi}} \frac{1}{\sqrt{1+p^2}}   \exp{\left(-i \xi
\arctan{( p)} \right)}  \quad .
\end{equation} 
For these states one has $\langle \hat p \rangle = 0, \langle \hat x
\rangle = \xi $ 
and \begin{equation}
\label{eqc19}
\langle \hat x \hat p +
\hat p \hat x \rangle = \int dp  \frac{1}{1+p^2} 
   \frac{2}{\pi} \left( \frac{- i p^2 + 2 \xi p + i}{1+p^2} \right) = 0 \quad ,
\end{equation}
so that Eq.(\ref{eqc14}) is verified. 

At this level, the natural question which arises from our analysis is the
existence (or not) of theories in which the term containing the anticommutator
may play a role in the uncertainty relation. This may happen if
 the only normalizable states obeying Eq.(\ref{eqc9})  correspond to  non  vanishing $\lambda_1$.
This excludes
models in which the commutation relations are of the form $[\hat x, \hat p] =
i f(\hat p)$  since in this case the relevant states read, in  momentum  space:
\begin{equation} \label{eqc20}
\psi(p) = N \exp{\left( - \int_{0}^{p} d\xi  \frac{(\lambda_2 \xi +
\lambda_4)}{f(\xi)} \right)}  \exp{\left( i \int_{0}^{p} d\xi 
\frac{(\lambda_1 \xi + \lambda_3)}{f(\xi)} \right)}  \quad .
\end{equation} 
Their  normalization requires the finiteness of the integral
\begin{equation}
\label{eqc21}
\langle \psi \vert \psi \rangle = N^2 \int_{-\infty}^{\infty}  \frac{dp}{f(p)} 
\exp{\left( -2 \int_{0}^{p} d\xi  \frac{(\lambda_2 \xi + \lambda_4)}{f(\xi)}
\right)} 
 \end{equation} 
which does not depend on $\lambda_1$. 

The models in which the deformation of the commutation relations is entirely
embodied in a function of the momentum are not the only ones which have
been studied in the literature. Among the 
proposals which do not fit in this category is for example the $q$
deformation of quantum mechanics  which is defined by the following equations 
\cite{wess}: 
\begin{equation}
\label{eqc22}
q^{\frac{1}{2}} \hat x \hat p - q^{-\frac{1}{2}} \hat p \hat x = i \hat
\Lambda  \quad , \quad \hat \Lambda \hat p = q \hat p \hat \Lambda  \quad ,
\quad  \hat \Lambda \hat x = q^{-1} \hat x \hat \Lambda  \quad .
\end{equation}
The argument we gave above can not be applied here. One needs a closer analysis to see if the new term
plays any role in such a model. We show in the next section that it proves to be 
important on the fuzzy sphere.

The reasoning followed here can be extended to higher dimensions. A partial 
motivation for such a treatment and an illustration in the three variables case is 
given in the appendix.

\section{Generalized Squeezed States On The Fuzzy Sphere.}

The saturation of the Heisenberg uncertainties (Eq.(\ref{eqb3})) 
related to the pairs of non commuting variables
translate into the formulas \cite{lubo1} 
\begin{eqnarray}
\label{eqd1}
\hat{m}_{jk} \vert \psi \rangle &=& 0 \quad {\rm with} \quad
\hat{m}_{12} = \hat x_1 + \mu_{12} \hat x_2 +  \tau_{12} \quad , \quad
\hat{m}_{23} = \hat x_2 + \mu_{23} \hat x_3 +  \tau_{23} \quad , \nonumber\\
\hat{m}_{31} &=& \hat x_3 + \mu_{31} \hat x_1 +  \tau_{31} \quad ,
\end{eqnarray}
where the $ \mu_{jk} $ are pure imaginary while the $ \tau_{jk}$ have real and
imaginary parts(see Eq.(\ref{eqc13})). We have adopted  simplified notations in this section; writing
$\mu_{jk}$ rather than $\bar{\mu}_{jk}$. No confusion is possible at this
stage, contrary to the previous section.  

Considering the following combinations of these equations
\begin{eqnarray}
\label{eqd2}
(\mu_{31} [ \hat{m}_{12}, \hat{m}_{23} ] +  [ \hat{m}_{23}, \hat{m}_{31} ]) \vert \psi \rangle 
&=& 0 \, , \,
(\mu_{23} [ \hat{m}_{31}, \hat{m}_{12} ] +  [ \hat{m}_{12}, \hat{m}_{23} ]) \vert \psi \rangle = 0 
\quad , \nonumber\\
(\mu_{12} [ \hat{m}_{23}, \hat{m}_{31} ] +   [ \hat{m}_{31}, \hat{m}_{12} ]) \vert \psi \rangle &=& 0 \, , \,
\end{eqnarray}
one obtains
\begin{equation}
\label{eqd3}
(1+ \mu_{12} \mu_{23} \mu_{31}) \hat x_1 \vert \psi \rangle = 0 \quad , \quad
(1+ \mu_{12} \mu_{23} \mu_{31}) \hat x_3 \vert \psi \rangle = 0 \quad , \quad
(1+ \mu_{12} \mu_{23} \mu_{31}) \hat x_2 \vert \psi \rangle = 0 \quad .
\end{equation}
Can the  three Heisenberg inequalities  be saturated  simultaneously? Only two cases
may lead to that situation:
\begin{itemize}
 \item The first possibility is
        \begin{equation}
        \label{eqd4}
          \hat x_k \vert \psi \rangle =  0 \quad , \quad k =1,2,3 \quad ,
        \end{equation}
     but then the second part of Eq.(\ref{eqb2}) is violated \quad .
 \item The remaining possibility
       \begin{equation}
       \label{eqd5}
          (1+ \mu_{12} \mu_{23} \mu_{31}) = 0
       \end{equation}
is ruled out by the fact that the quantities $\mu_{jk}$ are purely imaginary; it
 can be rewritten as
 \begin{equation}
 \label{eqd6}
 \frac{ \langle \hat x_1 \rangle \langle \hat x_2 \rangle  \langle \hat x_3 \rangle}
 {(\Delta x_1)^2 (\Delta x_2)^2 (\Delta x_3)^2 } = - 8 i \quad .
 \end{equation}
 \end{itemize}

Let us now turn to the enlarged uncertainty relations. Their saturation is displayed
in Eq.(\ref{eqb20}). The equation defining these "squeezed" states 
is the same as  Eq.(\ref{eqd1}) but now the coefficients
 $\mu_{jk}$ will have real and imaginary parts:
\begin{equation}
\label{eqd7}
\mu_{12} = \lambda_1 + i \lambda_2 \quad , \quad \mu_{23} = \lambda_3 + i 
\lambda_4 \quad , \quad
\mu_{31} = \lambda_5 + i \lambda_6 \quad \quad .
\end{equation}
The relation displayed in Eq.(\ref{eqd5}), which is necessary for the associated 
three uncertainties to be
saturated simultaneously, can be rewritten as
\begin{equation}
\label{eqd8}
1 + (\lambda_1 \lambda_3 - \lambda_2 \lambda_4) \lambda_5 -  (\lambda_1 \lambda_4 + \lambda_2 \lambda_3) 
\lambda_6 = 0
\quad , \quad
(\lambda_1 \lambda_3 - \lambda_2 \lambda_5) \lambda_6 +  (\lambda_1 \lambda_4 + \lambda_2 \lambda_3) \lambda_5 =
0 \quad .
\end{equation}
Not taking into account the terms $\tilde A_{x_i,x_j}$(defined in Eq.(\ref{eqc11})) amounts
 to impose  \newline
$\lambda_1=\lambda_3=\lambda_5= 0$
which transforms the first of the previous equations into the contradiction $1=0$.
The states we are looking for are in the kernels of the three operators 
$m_{jk}$ but with the 
coefficients $\mu_{j,k}$ having  non vanishing real parts.

For the remaining part of this work, we will adopt, for the fuzzy sphere, unitary representations of
dimensions $(2 j+1)(2 j+1)$:
\begin{eqnarray}
\label{eqd9}
 \hat{x}_k &=& \frac{1}{\sqrt{j(j+1)}} \hat{J}_k \quad , \quad
   \hat{J}_1 = \frac{1}{2} ( \hat{J}_+ + \hat{J}_-) \quad , \quad 
   \hat{J}_2 = -\frac{i}{2} ( \hat{J}_+ - \hat{J}_-) \quad , \nonumber\\
 \hat{J}_+ \vert m \rangle &=& \sqrt{(j - \mu)(j + m+1)} \vert m + 1 \rangle \quad , \quad
\hat{J}_- \vert m \rangle = \sqrt{(j + \mu)(j - m+1)} \vert m - 1 \rangle \quad , \nonumber\\
\hat{J}_3  \vert m \rangle &=& m \vert m \rangle \, .
\end{eqnarray}

We now solve Eq.(\ref{eqd1}) where the parameters $\mu_{jk}$ have real and imaginary 
parts. For a fixed $j$, the state we are looking for can be written as
\begin{equation}
\label{eqd10}
\vert Y \rangle = \sum_{n=-j}^j Y_n  \vert n \rangle  \quad .
\end{equation}
Each condition $\hat{m}_{jk} \vert Y \rangle = 0 $ leads to three "equations": the components
of $\vert -j \rangle, \vert j \rangle $ and $ \vert n \rangle$ for $-j+1 \leq n \leq j-1$.
For example,
\begin{eqnarray}
\label{ajout3}
\hat{m}_{12} \vert Y \rangle &=& \left[ \frac{1}{2} \alpha 
\left(  (1 - i \,\mu_{12}) \hat{J}_+ +  (1 + i \,\mu_{12}) \hat{J}_- \right) + \tau_{12} \right]
\vert Y \rangle \nonumber\\
&=& \frac{1}{2} \alpha   (1 - i \,\mu_{12})
\left( Y_{j-1} \sqrt{2 j} \, \vert j \rangle + \sum_{n=-j+1}^{j-1}
Y_{n-1} \sqrt{(j+n)(j-n+1)} \, \vert n \rangle  \right) \nonumber\\
&+& \frac{1}{2} \alpha   (1 + i \,\mu_{12})
\left( Y_{-j+1} \sqrt{2 j} \, \vert - j \rangle + \sum_{n=-j+1}^{j-1}
Y_{n+1} \sqrt{(j-n)(j+n+1)} \, \vert n \rangle  \right) \nonumber\\
&+& \tau_{12} \left( Y_{-j} \vert -j \rangle + 
Y_{j} \vert j \rangle + \sum_{n=-j+1}^{j-1} Y_{n} \vert n \rangle\right) \quad ,
\end{eqnarray}
where $\alpha=1/\sqrt{j(j+1)}$.
We end up with the following "nine" relations:
\begin{equation}
\label{eqd11}
\frac{Y_{-j+1}}{Y_{-j}} = - \frac{2 \tau_{12}}{ \alpha (1+ i \, \mu_{12} )
 \sqrt{2 j}}   \quad , 
\end{equation} 
\begin{equation}
\label{eqd12}
\frac{Y_{j}}{Y_{j-1}} = - \frac{ \alpha ( 1- i \, \mu_{12} ) \sqrt{2 j}}{2 
\tau_{12}}  \quad , 
\end{equation}
\begin{equation}
\label{eqd13}
Y_{n+1} +   \frac{2 \tau_{12}}{ \alpha (1+i\, \mu_{12} ) \sqrt{(j+n+1)(j-n)}}
 \, Y_n +
\, \frac{1- i\, \mu_{12}}{1+i \,\mu_{12}} \, 
\frac{\sqrt{(j-n+1)(j+n)}}{\sqrt{(j+n+1)(j-n)}}  \, Y_{n-1} = 0 \quad ,
\end{equation}
\begin{equation}
\label{eqd14}
\frac{Y_{-j+1}}{Y_{-j}} = 
\frac{2 (- \tau_{23} + \alpha \, \mu_{23} \, j)}{i\,\alpha \sqrt{2 j} } 
\quad , 
\end{equation}
\begin{equation}
\label{eqd15}
\frac{Y_{j}}{Y_{j-1}} = 
 \frac{i\,\alpha \, \sqrt{2 j}}{ 2 ( \alpha \, \mu_{23} j + \tau_{23} )}   \quad , \nonumber\\
\end{equation}
\begin{equation}
\label{eqd16} 
Y_{n+1} -   \frac{2 \,i\, (\alpha \, \mu_{23} n + \tau_{23} )}{ \alpha  \sqrt{(j+n+1)(j-n)}} 
 Y_n -
  \frac{\sqrt{(j-n+1)(j+n)}}{\sqrt{(j+n+1)(j-n)}}  \,  Y_{n-1} = 0 \quad ,
\end{equation}
\begin{equation}
\label{eqd17}
\frac{Y_{-j+1}}{Y_{-j}} = \frac{2 (\alpha \, j - \tau_{31})}{\alpha \, \mu_{31} \,
\sqrt{2 j} } 
\quad , 
\end{equation}
\begin{equation}
\label{eqd18}
\frac{Y_j}{Y_{j-1}} = -\frac{\alpha \, \mu_{31} \, \sqrt{2 j}}{ 2 
( \alpha \,  j + \tau_{31} )}   
\quad , 
\end{equation}
\begin{equation}
\label{eqd19}
Y_{n+1} + \frac{2  \, (\alpha \, n + \tau_{31} )}{ \alpha \, \mu_{31} \, \sqrt{(j+n+1)(j-n)}} 
 Y_n +
  \frac{\sqrt{(j-n+1)(j+n)}}{\sqrt{(j+n+1)(j-n)}}  \,  Y_{n-1} = 0 \quad .
\end{equation}

As this system has more equations than  unknowns, it will admit solutions only
when some relations between the 
coefficients hold. We now proceed to show how this happens.
Combining  Eq.(\ref{eqd11}) and Eq.(\ref{eqd14})  on one side, 
 Eq.(\ref{eqd11}) and Eq.(\ref{eqd17}) on the other side, one is
  led to the relations
\begin{equation}
\label{eqd20}
\tau_{23} = j \, \alpha \, \mu_{23} +  \frac{\tau_{12}}{-i+\mu_{12}}
\quad , \quad
\mu_{31} = - i \frac{(-i+\mu_{12}) (j \, \alpha - \tau_{31})}{\tau_{12}}
\quad .
\end{equation}
In the same way, from Eq.(\ref{eqd12}) , Eq.(\ref{eqd15}) ,  
and Eq.(\ref{eqd18}), one finds
\begin{equation}
\label{eqd21}
\mu_{23} = - i  \frac{\tau_{12}}{j \, \alpha \, ( 1+\mu_{12}^2 )}
\quad , \quad
\tau_{31} = - i \frac{j \alpha}{\mu_{12}} \quad .
\end{equation}
From these formula one finds the relation displayed in Eq.(\ref{eqd5})
which was shown to be necessary for the system to admit a solution.
Subtracting Eq.(\ref{eqd19}) from Eq.(\ref{eqd16}) , one has
\begin{equation}
 \label{eqd22}
 \sigma_n \equiv \frac{Y_{n}}{Y_{n-1}} =
 i\, j \, \sqrt{\frac{1+j-n}{j+n}}  \quad \frac{\alpha  
 (i+ \mu_{12})}{\tau_{12}} \quad .
\end{equation}
Using this result back in Eq.(\ref{eqd16}) one obtains 
\begin{equation}
\label{eqd23}
\tau_{12} =  j \, \alpha \, \sqrt{1+\mu_{12}^2} \quad .
\end{equation}
These expressions transform Eq.(\ref{eqd13}) into an identity.
So, the solution to our system of equations depends on the sole complex number
$\mu_{12}$; the other coefficients $\mu_{jk},\tau_{jk}$ as well as the
components of the state $\vert Y \rangle$. 

We  switch to a new variable $\zeta$ defined by the ratio of the last 
two components of our state:
\begin{equation}
\label{eqd24}
\frac{1}{\zeta} \equiv \frac{Y_{j}}{Y_{j-1}} = - \frac{1}{\sqrt{2 j}}  
\frac{1 -i\, \mu_{12}}
{\sqrt{1 + \mu_{12}^2}} \Longleftrightarrow  \mu_{12} = 
i \frac{2 j - \zeta^2}{2 j + \zeta^2}
\end{equation}
The coefficients of the set of equations now assume the form:
\begin{eqnarray}
\label{eqd25}
\mu_{23} &=&  \frac{i}{2 \sqrt{2 \, j}} \,\frac{2 j + \zeta^2}{ \zeta } \quad , \quad
\mu_{31} = - 2 \sqrt{2 \, j} \, \frac{ \zeta }{- 2 j + \zeta^2} \quad , \quad
\tau_{12} = -   \frac{ 2 \sqrt{2} \,j}{ \sqrt{1+j}} 
\frac{ \zeta }{ 2 j + \zeta^2} \quad , \nonumber\\
\tau_{23} &=& - \frac{i}{ 2 \, \sqrt{2} \, \sqrt{1+j}}  \,\frac{2 j - \zeta^2}{ \zeta } \quad , \quad
\tau_{31} = - \sqrt{ \frac{j}{j+1}} \, \frac{2 j + \zeta^2}{2 j - \zeta^2} \quad . 
\end{eqnarray}
Concerning the components of the state $\vert Y \rangle$, one has
\begin{eqnarray}
\label{eqd26}
\frac{Y_{-j+1}}{Y_{-j}} &=& \frac{2 j}{\zeta} \quad , \quad 
\frac{Y_{j}}{Y_{j-1}} = \frac{1}{\zeta} \quad , \quad
\frac{Y_{n}}{Y_{n-1}} = \sqrt{2 j} \,  \sqrt{\frac{1+j-n}{j+n}} 
\, \frac{1}{\zeta} \quad , \nonumber\\
& & {\rm when} -j+1 \leq n \leq j-1 \quad .
\end{eqnarray}
By a recursive reasoning, one can express all the components in terms
of the highest one
\begin{equation}
\label{ajout4}
Y_m = \frac{1}{(2 j)^{\frac{1}{2}(j-m)}} 
 \left( \begin{array}{c}
          2 j \\ j-m 
       \end{array}   \right)^{\frac{1}{2}}
\zeta^{j-m} \, Y_j \quad .
\end{equation}
The normalization of the wave function results in the condition
\begin{equation}
\label{ajout5}
Y_j^2 \left( \frac{\zeta \bar\zeta}{2 j} \right)^j   \sum_{m=-j}^{j}
\left( \begin{array}{c}
          2 j \\ j-m 
       \end{array}   \right)
\left( \frac{2 j}{\zeta \bar\zeta} \right)^m = 1  \quad .
\end{equation}
Rewriting  this sum in terms of the index $n=m+j$ which goes from $0$ to $2 j$, one obtains
\begin{equation}
\label{ajout6}
Y_j =  \left( \frac{2 j}{2 j+ \zeta \bar\zeta} \right)^j  \quad .
\end{equation}
Replacing this in the expression of $Y_m$ and denoting from now on the state
by the complex parameter $\zeta$, one obtains
\begin{equation}
\label{eqd27}
\vert \zeta \rangle = \frac{1}{(2 j+ \zeta \bar\zeta)^j} \sum_{m=-j}^j  \, R_m \, \zeta^{-m+j} 
\, \vert m \rangle \quad ,
\end{equation}
 where the real constants $R_m$ are given by
\begin{eqnarray}
\label{eqd28}
R_m &=&  (2 j)^{\frac{1}{2}(m+j)} 
 \left( \begin{array}{c}
          2 j \\ j-m 
       \end{array}   \right)^{1/2}
 {\rm for \quad all} \quad  n  \quad {\rm since} \quad  \nonumber\\
R_{-j} &=&  1 \quad {\rm and} \quad R_{j} = (2 j)^{j} \quad .
\end{eqnarray}

We see that considering  the uncertainty relation we derived in the third section, 
which is weaker than Heisenberg's,
we pass from a situation in which no state saturates all the paired bounds to
one in which there is a family of states achieving that. 
This family  is parameterized by one complex number $\zeta$,
like the generalized coherent states on the fuzzy sphere obtained in 
\cite{pinzul1}. The difference between the two works is twofold. Firstly, the starting
point here is the uncertainty relation. Secondly, the Fock space in which we work
is finite dimensional so that we don't need a truncature procedure. 

After some algebra, one finds for the scalar product the expression
\begin{equation}
\label{eqd29}
\langle \zeta  \vert \eta \rangle = \frac{(2 j+ \bar\zeta \eta)^{2 j}}
{(2 j+ \bar\zeta \zeta)^{j} (2 j+ \bar\eta \eta)^{j}} \quad .
\end{equation}
Using the formula
\begin{equation}
\label{eqd30}
\lim_{x \rightarrow \infty} \left( 1 + \frac{a}{x} \right)^x = e^a  \quad ,
\end{equation}
one obtains in the limit $j \rightarrow \infty$ the formula
\begin{equation}
\label{eqd31}
\vert \langle \eta  \vert \zeta \rangle \vert^2 = e^{- \vert \eta - \zeta \vert^2 } \quad .
\end{equation}
which is also valid for the ordinary coherent states on the non commutative 
plane. It has been known that there is a limit in  the parameter space of the 
fuzzy sphere which reproduces the non commutative plane \cite{madore3}.
The result obtained here goes in the same direction: the two structures have
things in common.

However, contrary to the non commutative plane, for any fixed generalized squeezed 
state  $ \vert \zeta \rangle $,
there is another 
state $ \vert \eta \rangle $ to which
it is orthogonal; it satisfies $ \zeta \bar\eta = - 2 j$. However, it disappears in
the commutative limit.

\section{The Stereographic Projection.}

We now show how the states built so far reproduce the stereographic projection.
The mean value $\langle x_1\rangle$ can be written as
\begin{eqnarray}
\label{eqe1}
\langle \zeta \vert \hat{x}_1 \vert \zeta\rangle &=& \frac{1}{\sqrt{j(j+1)}}
\frac{\zeta^j {\bar\zeta}^j }{(2 j+\zeta \bar\zeta)^{2j}}
\sum_{m,n} R_m R_n \zeta^{-m} {\bar\zeta}^{-n}
\langle n \vert J_1 \vert m \rangle
\end{eqnarray} 
Expressing $J_1$ as in Eq.(\ref{eqd9}), we can split the sum in the previous
equation 
\begin{equation}
\label{eqe2}
\sum_{m,n} R_m R_n \zeta^{-m} {\bar\zeta}^{-n}
\langle n \vert J_1 \vert m \rangle = I_1 + I_2 \quad .
\end{equation}
The first term can be computed as follows
\begin{eqnarray}
\label{eqe3}
I_1 &=&  \sum_{m=-j}^{j-1} R_m R_{m+1} \zeta^{-m} {\bar\zeta}^{-m-1} 
\sqrt{(j+m+1)(j-m)} \nonumber\\ 
&=& \zeta \sum_{m=-j}^{j-1}    (2 j)^{m+j+3/2} 
\left( \begin{array}{c}
          2 j -1\\ j+m 
       \end{array}   \right) (\zeta \bar\zeta)^{-m-1} \nonumber\\
&=& (2 j)^{j+3/2} \, \frac{1}{\bar\zeta} \, \sum_{n=0}^{2 j-1}     
\left( \begin{array}{c}
          2 j -1\\ n 
       \end{array}   \right)  \,
       \left( \frac{2 j}{\zeta \bar\zeta} \right)^{n-j} \nonumber\\
&=& (2 j)^{3/2} \, \zeta \, \frac{(2 j+ \zeta \bar\zeta)^{2 j-1}}
{(\zeta \bar\zeta)^j} \quad .
\end{eqnarray}
Similarly,
\begin{eqnarray}
\label{eqe4}
I_2 &=&  \sum_{m=-j+1}^{j} R_m R_{m-1} \zeta^{-m} {\bar\zeta}^{-m+1} 
\sqrt{(j-m+1)(j+m)}   \nonumber\\
&=& 
(2 j)^{3/2} \, \bar\zeta \, \frac{(2 j+ \zeta \bar\zeta)^{2 j-1}}
{(\zeta \bar\zeta)^j}  \quad ,
\end{eqnarray}
so that the wanted mean value rakes the form
\begin{equation}
\label{eqe5}
\langle \zeta \vert \hat{x}_1 \vert \zeta\rangle =  \frac{\sqrt{2} j}{\sqrt{j+1}}
\frac{\zeta + \bar\zeta}{2 j+ \zeta \bar\zeta} \quad .
\end{equation}
The two remaining quantities $\langle \hat{x}_k\rangle$ can be obtained 
using the fact that Eq.(\ref{eqd1}) implies $\langle \hat{m}_{jk}\rangle= 0 .$
For example, the simplification
\begin{eqnarray}
\label{eqe6}
\tau_{12} + \langle \zeta \vert \hat{x}_1 \vert \zeta\rangle &=& \frac{\sqrt{2} j}{\sqrt{j+1}}
\left(- \frac{2 \zeta}{2 j+\zeta^2} + 
\frac{\zeta + \bar\zeta}{2 j+ \zeta \bar\zeta}\right) \nonumber\\
&=& \frac{\sqrt{2} j}{\sqrt{j+1}} 
\frac{(\bar\zeta - \zeta)(2 j-\zeta^2)}{(2 j+\zeta^2)(2 j+ \zeta \bar\zeta)} 
\end{eqnarray}
allows one to infer from $\langle \hat{m}_{12}\rangle= 0$ the mean value
\begin{equation}
\label{eqe7}
\langle \zeta \vert \hat{x}_2 \vert \zeta\rangle = 
- \frac{1}{\mu_{12}} (\tau_{12} + \langle \zeta \vert \hat{x}_1 \vert \zeta\rangle)
= i \, \frac{\sqrt{2} j}{\sqrt{j+1}}
\frac{\bar\zeta - \zeta}{2 j+ \zeta \bar\zeta} \quad .
\end{equation}
In the same way,
\begin{equation}
\label{eqe8}
\langle \zeta \vert \hat{x}_3 \vert \zeta\rangle = 
- \left( \langle \zeta \vert \hat{x}_1 \vert \zeta\rangle + 
\frac{\tau_{31}}{\mu_{31}} \right) \mu_{31} =
 - \sqrt{\frac{j}{j+1}}
\frac{-2 j+ \zeta \bar\zeta}{2 j+ \zeta \bar\zeta} \quad ,
\end{equation}
while the vanishing of  $\langle \hat{m}_{23} \rangle$ results in an identity.

The expressions of these mean values resemble the ones defining the stereographic
projection. The link can be made more transparent if one switches to the
variable 
\begin{equation}
\label{eqe9}
\zeta = \sqrt{2 j} \beta \quad 
\end{equation}
since the mean values now read
\begin{equation}
\label{eqe10}
\langle \hat{x}_1 \rangle = \sqrt{\frac{j}{j+1}} \,
\frac{\beta + \bar\beta}{1 + \beta  \bar\beta} \quad , \quad
\langle \hat{x}_2 \rangle = i\, \sqrt{\frac{j}{j+1}} \,
\frac{\bar\beta - \beta}{1 + \beta  \bar\beta} \quad , \quad
\langle \hat{x}_3 \rangle = \sqrt{\frac{j}{j+1}} \,
\frac{ \beta  \bar\beta - 1}{1 + \beta  \bar\beta} \quad .
\end{equation}
The stereographic map is then recovered in the large $j$ limit.
Concerning the dispersions, one has
\begin{eqnarray}
\label{ajout7}
1 &=& \langle \hat{x}_1^2 \rangle + \langle \hat{x}_2^2 \rangle + 
\langle \hat{x}_3^2 \rangle 
\quad , \quad
\frac{j}{j+1} = \langle \hat{x}_1 \rangle^2 + \langle \hat{x}_2 \rangle^2 + 
\langle \hat{x}_3 \rangle^2 \quad , \nonumber\\
&\Longrightarrow & (\Delta x_1)^2 + (\Delta x_2)^2 + (\Delta x_3)^2 = \frac{1}{j+1}
\end{eqnarray}
so that in the large $j$ limit all the three $\Delta x_k$ vanish.

In terms of this new variable, the generalized squeezed state reads
\begin{equation}
\label{eqe11}
\vert \beta \rangle = 
 \left(\frac{1}{1+ \beta \bar\beta} \right)^j 
 \sum_{m=-j}^{j} 
 \left( \begin{array}{c}
          2 j \\ j -m 
       \end{array}   \right)^{1/2}
  \beta^{-m+j} 
 \vert m \rangle \quad ,
\end{equation}
while the scalar product takes the form
\begin{equation}
\label{ajout8}
\langle \gamma \vert \beta \rangle = 
\left( \frac{(1+ \bar \gamma \beta)^2}{(1+ \gamma \bar\gamma)(1+ \beta \bar\beta)}
\right)^j  \quad .
\end{equation}

We can compute these mean values in a different way. For example,
from Eq.(\ref{eqc7}) we have
\begin{equation}
\label{ajout9}
\langle \hat p \rangle = -\frac{\bar{\lambda}_4}{\bar{\lambda}_2} \quad .
\end{equation}
Considering the equation $ \hat{m}_{12} \vert \zeta \rangle = 0$, we have to replace
$\hat x$ by $\hat{x}_1$  and $\hat p$ by $\hat{x}_2$. The separation between
real and imaginary parts
$\mu_{12} = \bar{\lambda}_{1}+ i \bar{\lambda}_2 \quad , \quad
\tau_{12} = \bar{\lambda}_{3}+ i \bar{\lambda}_4 $ 
reads
\begin{equation}
\label{ajout10}
\bar{\lambda}_2 = \frac{(2 j+ \zeta \bar\zeta)(2 j- \zeta \bar\zeta)}
{(2 j+ \zeta^2)(2 j+ {\bar\zeta}^2)} 
\quad , \quad
\bar{\lambda}_4 = - \frac{2 \sqrt{2} j}{\sqrt{1+j^2}} \frac{\zeta_2 (2 j- \zeta \bar\zeta)}
{(2 j+ \zeta^2)(2 j+ {\bar\zeta}^2)} \quad ,
\end{equation}
$\zeta_2$ being the imaginary part of $\zeta$. Replacing this into Eq.(\ref{ajout9})
appropriately modified, one recovers Eq.(\ref{eqe7}). 

We shall finish this section with the simplest possible illustration, i.e the 
case $j=1/2$. The variables $\zeta$ and $\beta$ then coincide.
The coefficients of the operators $\hat{m}_{jk}$ have the form
\begin{eqnarray}
\label{eqe13}
\mu_{1,2} &=&  i \frac{1 - \zeta^2}{1 + \zeta^2} \quad , \quad 
\tau_{1,2} = - \frac{2}{\sqrt{3}} \frac{\zeta}{1+ \zeta^2} \quad , \quad
\mu_{2,3} = i \frac{1+\zeta^2}{2 \zeta} \quad , \nonumber\\
\tau_{2,3} &=& - i \frac{1}{2 \sqrt{3}} \frac{1 -\zeta^2}{\zeta} \quad , \quad
\mu_{3,1} = \frac{2 \zeta}{1-\zeta^2} \quad , \quad
\tau_{3,1} = - \frac{1}{\sqrt{3}} \frac{1+\zeta^2}{1-\zeta^2} \quad ,
\end{eqnarray} 
while the generalized squeezed states read
\begin{eqnarray}
\label{eqe14}
\vert \zeta \rangle = \frac{1}{\sqrt{1 + \vert \zeta \vert^2}}  
\left( \zeta \vert - 1/2 \rangle + \vert  1/2 \rangle \right) \quad .
\end{eqnarray}
The deformed stereographic projection takes the form
\begin{equation}
\label{eqe15}
\langle \hat{x}_1 \rangle = \frac{1}{\sqrt{3}} \frac{\bar\zeta + \zeta}
{1 +\zeta \bar\zeta } \quad , \quad
\langle \hat{x}_2 \rangle = \frac{i}{\sqrt{3}} \frac{\bar\zeta - \zeta}
{1 +\zeta \bar\zeta } \quad , \quad
\langle \hat{x}_3 \rangle = \frac{1}{\sqrt{3}} \frac{1- \bar\zeta  \zeta}
{1 +\zeta \bar\zeta } \quad .
\end{equation}

\section{Conclusions.}
  We have derived an uncertainty relation which is weaker than the Heisenberg's.
On the fuzzy sphere, this relation can be saturated for all the pairs of variables.
The states built in this way possess some interesting properties. They are 
parameterized by a complex number and realize the stereographic map. This fact 
is promising in the sense that the star product associated to these states
will naturally be defined on functions of two real variables i.e exactly the
geometric dimension of the sphere \cite{lubo2}.

The link between our states and the geometry of the sphere suggest the
possibility of studying Q.F.T in this context using a star product
defined on the stereographic coordinates. An important question will then be 
the properties of field theory on the fuzzy sphere \cite{klimcik} which are 
recovered in such a formulation. A study of classical solutions \cite{scho,landi} 
would of course
be interesting.

\appendix

\section{{\bf Appendix}: High Dimensional  Extension.}

There exists a model possessing a minimal uncertainty in length
while preserving rotational and translational symmetries 
\cite{kmm1}. Its
non trivial commutation relations are:
\begin{equation}
\label{eqa2}
[ \hat x_j , \hat p_k] = i \hbar \left(  f(\hat{{\vec p}^2})  \delta_{j
k} + g(\hat{{\vec p}^2}) \hat p_j  \hat p_k \right) \quad , \end{equation} 
supplemented by the condition 
\begin{equation}
\label{eqa3}
g = \frac{2 f f'}{f - 2 p^2 f'} \quad .
\end{equation} 
The functions $f$ and $g$ are supposed positive. This theory admits four  
couples of non commuting variables \cite{lubo1}:
$(x_1,p_1),(x_2,p_2),(x_1,p_2)$ and $(x_2,p_1)$. It was shown in \cite{lubo1}
that one can not saturate the associated non trivial uncertainties simultaneously.
The same argument can be repeated here to show this is still the case even for the
enlarged uncertainties we derived. How can one define squeezed states
which do not discriminate between the variables? Although we will not apply it
explicitly, we think one may be led to look at the structure of uncertainty
relations implying more than two operators. We outline this below. 

In the second section, we have seen that in $2 D$ an analysis based
on an operator leads to an uncertainty implying
the mean values of at most quadratic expressions of the variables. The question we
would like to answer  in this section is the following: what are
the  general uncertainties implying more than two variables? 

We now go one step further and consider three variables simultaneously. Mimicking what was done in the second section,
we  consider an operator
of the form
 \begin{equation}
\label{eqa4}
\hat{\Theta}_\lambda = \hat x_1 + ( \lambda_1 + i \lambda_2) \hat x_2 + ( \lambda_3 +
i \lambda_4) \hat p_1  +  ( \lambda_5 + i \lambda_6) I  \quad .
\end{equation}
We  build the function $\varphi(\vec\lambda)$ similarly to what 
was done in Eq.(\ref{eqc3}):
\begin{eqnarray}
\label{eqa5}
\varphi(\lambda_1, \lambda_2,\lambda_3, \lambda_4,\lambda_5, \lambda_6) &=& 
\langle \hat{x}_1^2 \rangle + \lambda_1 A_{x_1,x_2} +
 \lambda_2 C_{x_1,x_2} + 
\lambda_3 A_{x1,p1} + \lambda_4 C_{x_1,p_1}  + 2 \lambda_5 \langle \hat{x}_1 
\rangle  \nonumber\\
&+& (\lambda_1^2 + \lambda_2^2) \langle \hat{x}_2^2 \rangle +
(\lambda_1 \lambda_3 + \lambda_2 \lambda_4) A_{x_2,p_1}+ 
(\lambda_1 \lambda_4 - \lambda_2 \lambda_3) C_{x_2,p_1} \nonumber\\
&+& 2 (\lambda_1 \lambda_5 + \lambda_2 \lambda_6) \langle  \hat{x}_2 \rangle 
+ (\lambda_3^2 + \lambda_4^2)  \langle \hat{p}_1^2  \rangle+
2 (\lambda_3 \lambda_5 + \lambda_4 \lambda_6) \langle  \hat{p}_1 \rangle \nonumber\\
&+&  
(\lambda_5^2 + \lambda_6^2)
\end{eqnarray}
We can diagonalize it, similarly to what was done in Eq.(\ref{eqc5}):
\begin{eqnarray}
\label{eqa6}
\varphi( \lambda_1, \lambda_2,\lambda_3, \lambda_4,\lambda_5, \lambda_6) &=&  
\sigma_1 \, ( \lambda_1 + \mu_{12} \, \lambda_2 + \mu_{13} \, \lambda_3+ 
 \mu_{14} \, \lambda_4 + \mu_{15} \, \lambda_5+ \mu_{16} \, \lambda_6 +
  \mu_{10})^2 \nonumber\\
  &+& \sigma_2 \, ( \quad \quad \quad \quad  \lambda_2 + \mu_{23} \, \lambda_3+ 
 \mu_{24} \, \lambda_4 + \mu_{25} \, \lambda_5+ \mu_{26} \, \lambda_6 +
  \mu_{20})^2 \nonumber\\
  &+& \sigma_3 \, ( \quad \quad \quad \quad \quad \quad \quad \quad \lambda_3 + 
  \mu_{34} \, \lambda_4 + 
  \mu_{35} \, \lambda_5+ \mu_{36} \, \lambda_6 + 
  \mu_{30})^2 \nonumber\\
   &+& \sigma_4 \, (\quad \quad \quad \quad \quad \quad \quad \quad \quad \quad \quad \quad   \lambda_4 + 
  \mu_{45} \, \lambda_5+ \mu_{46} \, \lambda_6 + 
  \mu_{40})^2 \nonumber\\
  &+& \sigma_5 \, (\quad \quad \quad \quad \quad \quad \quad \quad \quad \quad \quad \quad \quad \quad \quad \quad
   \lambda_5+ \mu_{56} \, \lambda_6 + 
   \mu_{50})^2 \nonumber\\
   &+& \sigma_6 \, (\quad \quad \quad \quad\quad \quad \quad \quad \quad \quad \quad \quad \quad \quad \quad 
   \quad \quad \quad \quad \quad
   \lambda_6 + 
   \mu_{60})^2 \nonumber\\
   &+& \sigma_7 \quad .
\end{eqnarray}

The coefficients $\sigma_i$ and $\mu_{j,k}$(which have nothing to do with those
of section $4$) are easily found:
\begin{eqnarray}
\label{eqa7}
\sigma_1 &=& \langle \hat{x}_2^2 \rangle \quad , \quad \mu_{12}= 0 \quad , \quad
\mu_{13}= \frac{A_{x_2,p_1}}{2 \langle  \hat{x}_2^2 \rangle} \quad ,  \quad
\mu_{14}= \frac{C_{x_2,p_1}}{2 \langle  \hat{x}_2^2 \rangle}  
 , \quad
\mu_{15}= \frac{\langle \hat{x}_2 \rangle}{\langle  \hat{x}_2^2 \rangle} 
\quad ,  \nonumber\\
\mu_{16} &=& 0 \quad , \quad \mu_{10} = 
\frac{A_{x_1,x_2}}{2 \langle  \hat{x}_2^2 \rangle} \quad ,  
\end{eqnarray}
\begin{eqnarray}
\label{eqa8}   
\sigma_2 &=& \sigma_1 \quad , \quad \mu_{23} = - \mu_{14} \quad ,
\quad \mu_{24} =  \mu_{13} \quad , \quad
 \mu_{25} = 0 \quad , \quad \mu_{26} =  \mu_{15} \quad ,
 \nonumber\\ 
\mu_{20} &=& \frac{C_{x_1,x_2}}{2 \langle  \hat{x}_2^2 \rangle} \quad , 
\end{eqnarray}
\begin{eqnarray}
\label{eqa9}
\sigma_3 &=&\frac{ - A_{x_2,p_1}^2 - C_{x_2,p_1}^2 + 4 \langle \hat{p}_1^2 \rangle \langle \hat{x}_2^2 \rangle}
{4 \langle \hat{x}_2^2 \rangle} \quad , \quad \mu_{34} = 0 \quad , 
\nonumber\\ 
\mu_{35} &=& \frac{2 ( - 2 \langle \hat{p}_1 \rangle  \langle \hat{x}_2^2 \rangle  + 
A_{x_2,p_1} \langle \hat{x}_2 \rangle )}{A_{x_2,p_1}^2 + C_{x_2,p_1}^2 - 4 \langle \hat{p}_1^2 
\rangle \langle \hat{x}_2^2 \rangle} \quad , \quad
\mu_{36}  =  -\frac{2\, C_{x2,p1} \, \langle \hat{x}_2 \rangle}
  { A_{x2,p1}^2 + C_{x2,p1}^2 - 
    4\, \langle  \hat{p}_1^2 \rangle \langle  \hat{x}_2^2 \rangle} \quad , 
    \nonumber\\
\mu_{30} &=& \frac{ A_{x1,x2} A_{x2,p1} - 
    C_{x1,x2} C_{x2,p1} - 
    2\, A_{x1,p1} \, \langle  \hat{x}_2^2 \rangle }{ A_{x2,p1}^2 + C_{x2,p1}^2 - 
    4\, \langle  \hat{p}_1^2 \rangle , \langle  \hat{x}_2^2 \rangle}  \quad , \nonumber\\  
\end{eqnarray}
\begin{eqnarray}
\label{eqa10}
\sigma_4 &=& \sigma_3 \quad , \quad \mu_{45} = - \mu_{36} \quad  , \quad \mu_{46} =  \mu_{35} \quad , 
\nonumber\\
\mu_{40} &=& \frac{ A_{x2,p1} C_{x1,x2} + 
    A_{x1,x2} C_{x2,p1} - 
    2\, C_{x1,p1}   \, \langle \hat{x}_2^2 \rangle
    }{ A_{x2,p1}^2 + C_{x2,p1}^2 - 
    4\, \langle \hat{p}_1^2 \rangle \langle \hat{x}_2^2 \rangle } \quad , \nonumber\\
\end{eqnarray}
\begin{eqnarray}
\label{eqa11}
\sigma_5 &=& 1 - \mu_{15}^2 \, \sigma_1 - ( \mu_{35}^2 \, +  \mu_{45}^2 ) \, \sigma_3  \quad , \quad
\mu_{56}= 0  \quad 
\nonumber\\
\mu_{50} &=&  \frac{1}{\sigma_5} 
\left[ \langle \hat{x}_1 \rangle  - \mu_{10} \, \mu_{15} \, \sigma_1 -( \mu_{30}  \, \mu_{35} - 
\mu_{36} \, \mu_{40}) \, \sigma_3 \right] \quad , \nonumber\\
\end{eqnarray}
\begin{eqnarray}
\label{eqa12}
\sigma_6 &=&   \sigma_5 \quad , \quad \mu_{60} =  - \frac{1}{\sigma_5} \left[ \mu_{15} \mu_{20} \sigma_1 + 
(\mu_{36} \mu_{30} + \mu_{35} \mu_{40}) \sigma_3 \right] 
 \quad , \nonumber\\
\end{eqnarray}
\begin{eqnarray}
\label{eqa13}
\sigma_7 &=&  \langle \hat{x}_1^2 \rangle - (\mu_{10}^2  + \mu_{20}^2 ) \sigma_1 - 
( \mu_{30}^2  +  \mu_{40}^2 ) \sigma_3 - ( \mu_{50}^2  +  \mu_{60}^2 ) \sigma_5  \quad .
\end{eqnarray}

As in the previous section, one has the inequalities
\begin{eqnarray}
\label{eqa14}
 \sigma_1,...,\sigma_6 > 0 \quad , \quad \sigma_7 \geq 0 \quad .
\end{eqnarray}

 For a fixed state $\vert \psi \rangle$, 
the minimum of the function $\varphi$ is attained for special values 
 $ \bar{\lambda}_1,...,\bar{\lambda}_6$ which can be expressed in function of the observables of the 
 states. For example, 
 \begin{eqnarray}
\label{eqa15}
\bar{\lambda}_6 &=& - \mu_{60} \quad , \quad \bar{\lambda}_5 = - \mu_{50}  - 
\mu_{56} \bar{\lambda}_6  \quad , \quad
\bar{\lambda}_4 =  - \mu_{40}  - 
\mu_{45} \bar{\lambda}_5  - 
\mu_{46} \bar{\lambda}_6 \quad , 
\end{eqnarray} 
where  the quantities $\mu_{jk}$ are given in 
Eqs.(\ref{eqa7},\ref{eqa8},\ref{eqa9},\ref{eqa10},\ref{eqa11},\ref{eqa12},\ref{eqa13}).

We are now in a position to characterize the kernel of the operator 
$\hat{\Theta}_{\bar\lambda}$
of 
Eq.(\ref{eqa4}). For such elements, the function $\varphi$ attains its minimum
so that the parameters of the operator are linked to observables by the relations
displayed above. As is equals zero, one has in addition that
 \begin{eqnarray}
\label{eqa16}
\sigma_7 = 0 \quad .
\end{eqnarray} 
We have obtained an uncertainty mixing  the three coordinates and which can be 
saturated.

\end{document}